\documentclass[a4paper]{article}
\usepackage{INTERSPEECH2021}
\usepackage[caption = false]{subfig}
\usepackage{comment}
\usepackage{cite}
\usepackage{todonotes}

\newcommand{\zvi}[1]{\mathbf{z}_{#1}}
\newcommand{\svi}[1]{\mathbf{s}_{#1}}
\newcommand{\cvi}[1]{\mathbf{c}_{#1}}

\title{Segmental Contrastive Predictive Coding for Unsupervised Word Segmentation}

\name{Saurabhchand Bhati$^{\dagger}$, Jes\'us Villalba$^{\dagger,\ddagger}$, Piotr \.Zelasko$^{\dagger,\ddagger}$, Laureano Moro-Velazquez$^{\dagger}$, Najim Dehak$^{\dagger,\ddagger}$}
\address{$^{\dagger}$Center for Language and Speech Processing,
 $^{\ddagger}$Human Language Technology Center of Excellence,  \\ Johns Hopkins University, Baltimore, MD, USA}

\email{\{sbhati1,jvillalba,pzelasko,laureano,ndehak3\}@jhu.edu}

\begin{document}
\maketitle

\begin{abstract}
Automatic detection of phoneme or word-like units is one of the core objectives in zero-resource speech processing. Recent attempts employ self-supervised training methods, such as contrastive predictive coding (CPC), where the next frame is predicted given past context. However, CPC only looks at the audio signal's frame-level structure. We overcome this limitation with a segmental contrastive predictive coding (SCPC) framework that can model the signal structure at a higher level e.g. at the phoneme level. In this framework, a convolutional neural network learns frame-level representation from the raw waveform via noise-contrastive estimation (NCE). A differentiable boundary detector finds variable-length segments, which are then used to optimize a segment encoder via NCE to learn segment representations. The differentiable boundary detector allows us to train frame-level and segment-level encoders jointly. Typically, phoneme and word segmentation are treated as separate tasks. We unify them and experimentally show that our single model outperforms existing phoneme and word segmentation methods on TIMIT and Buckeye datasets. We analyze the impact of boundary threshold and when is the right time to include the segmental loss in the learning process.
\end{abstract}

\section{Introduction}

Unsupervised discovery of phoneme or word-like units forms the core objective of Zero Resource speech processing~\cite{jansen2011efficient,badino2014auto,huijbregts2011unsupervised,lee2012nonparametric,siu2014unsupervised,kamper2017segmental,bhati2017unsupervised,kamper2017embedded,bhati2018phoneme}. The representation techniques employed to characterize the speech signal dictate the quality of the segmentation and clustering and, in turn, the entire process of discovering linguistic units from the speech signal. 
A good speech representation becomes crucial for unsupervised systems' good performance. 
Self-supervised methods have emerged as a promising technique for representation learning from unlabeled speech data~\cite{oord2018representation,schneider2019wav2vec,baevski2020wav2vec,kreuk2020self}.  
In the self-supervised learning (SSL) scenario, the `self' part refers to the generation of pseudo-labeled training data for an auxiliary task, and `supervised' refers to the supervised training of the underlying neural model. 
While most SSL work in speech focuses on learning representations~\cite{oord2018representation,schneider2019wav2vec,baevski2020wav2vec}, SSL has recently been used to identify the spectral changes from the raw waveform and detect phoneme boundaries~\cite{kreuk2020self}. 

The SSL framework exploits the temporal structure present in the speech to learn latent representations. In~\cite{kreuk2020self}, the auxiliary task is to identify the next frame's latent representation given the latent representation of a reference frame. A feature extractor, e.g., Convolutional Neural Network (CNN), that maps the speech signal to a latent space is optimized using the Noise Contrastive Estimation (NCE)~\cite{gutmann2010noise} to correctly identify the next frame within a set of frames that includes random distractor frames. 
The speech signal is composed of underlying linguistic units like phonemes or words. Thus it would be beneficial to learn from the speech structure at phoneme level. But phoneme segmentation is not straightforward and phoneme length is variable, which makes working at phoneme level difficult. That is why current SSL frameworks remain limited to structure at the frame level. 

In this work, we present a new method for unsupervised phoneme and word segmentation, for which we propose Segmental Contrastive Predictive Coding (SCPC). This framework can exploit the structure in speech signal at a high scale, i.e., phonemes, during the learning process. A word can be thought of as a sequence of segments. For a certain language, and considering the phonemic domain, segments in a word will occur together frequently and will have a higher likelihood of following each other than segments across words. By locating lower prediction probability time points between adjacent segments, we can get the word boundary candidates.

In the proposed approach, we start from the raw waveform and train an encoder via the next frame prediction to extract frame-level latent representations. We then use a differentiable boundary detector to extract variable-length segments. We develop a simple yet efficient differentiable boundary detection that can detect boundaries without any constraints on segment length or the number of segments in an utterance. Then, we obtain the segment level representation by encoding the averages of the variable-length segments through a segment encoder. The differentiable boundary detector allows information to flow between frame and segment encoder. The model is trained in an end-to-end manner. 

Our proposed methods enable boundary detection and segment representation computation in a batch manner for faster training. At the frame level, the model is optimized to predict the next frame. At the segment level, the auxiliary task becomes the next segment prediction.
Unlike frames, for segments, the previous segment might not be enough to predict the next segment. We need to learn the context or the order in which the segments occur in. We use a Recurrent neural network to capture the segment context.
The joint training allows the model to capture the structure present in the speech at multiple levels, i.e., frames and phonemes level, and these two mutually benefit from each other. We evaluate our proposed methods on TIMIT~\cite{garofolo1993timit} and Buckeye~\cite{pitt2005buckeye} datasets for phoneme and word boundary detection. Our proposed method outperforms state-of-the-art phoneme and word segmentation methods.

\section{Related Work }
Most previous works reduce the phoneme boundary detection task to a boundary classification task at each time step.
Michel et al.~\cite{michel2016blind} use HMM or RNN to predict the subsequent frame. A peak detection algorithm identifies the high prediction error regions, which are then used as phoneme boundaries. Wang et al.~\cite{wang2017gate} train an RNN autoencoder and track the norm of intermediate gate values over time e.g., forget gate for LSTM. A peak detection algorithm identified the boundaries from the gate norm over time.  Kreuk et al.~\cite{kreuk2020self} train a CNN encoder to distinguish between adjacent frame pairs and random pair of distractor frames. To detect the phoneme boundaries, a peak detection algorithm is applied over the model outputs. This method achieves state-of-the-art phoneme segmentation on TIMIT and Buckeye dataset. All these methods try to exploit the structure at frame level to detect phoneme boundaries. 

Word segmentation is an important problem in Zero resource speech processing. Bayesian Segmental GMM~\cite{kamper2017segmental} and Embedded Seegmental K-Means~\cite{kamper2017embedded} both start from an initial set of subword boundaries and then iteratively eliminate some of the boundaries to arrive at frequently occurring longer word patterns. The initial subword boundaries are not adjusted during the process. As a result, the performance of the ES-Kmeans critically depends on the initial boundaries~\cite{bhati2018phoneme}.  Kamper et al.~\cite{kamper2020towards} proposed methods to constrain the outputs of vector-quantized~(VQ) neural networks and assign blocks of consecutive feature vectors to the same segment. 
First, vector quantized variational autoencoder~(VQ-VAE)~\cite{chorowski2019unsupervised} and vector quantized contrastive predictive coding~(VQ-CPC)~\cite{baevski2019vq} models are trained to encode the speech signal into discrete latent space. Then, dynamic programming~(DP) is used to merge the frames to optimize a squared error with a length penalty term to encourage longer but fewer segments. The DP segmentation output is further segmented using word segmentation algorithms, e.g. adopter grammar~\cite{johnson2007adaptor}, or the Dirichlet process model~\cite{goldwater2009bayesian},  typically used for segmenting sequences of space-removed characters or phonemes. 

The tasks of phoneme segmentation and word segmentation are often done via different methods~\cite{kamper2017embedded,kamper2017segmental,bhati2018phoneme,bhati2019unsupervised}. Some methods use phoneme segmentation as a starting point for word boundaries~\cite{bhati2018phoneme,bhati2019unsupervised}. However, none of the methods jointly do phoneme and word segmentation. Here, we propose a single system capable of doing both and able to exploit the two tasks' inter-dependencies. 

\section{Segmental Contrastive Predictive Coding}

\begin{figure}
    \hspace*{-0.04\linewidth}
    \includegraphics[width=3.3in]{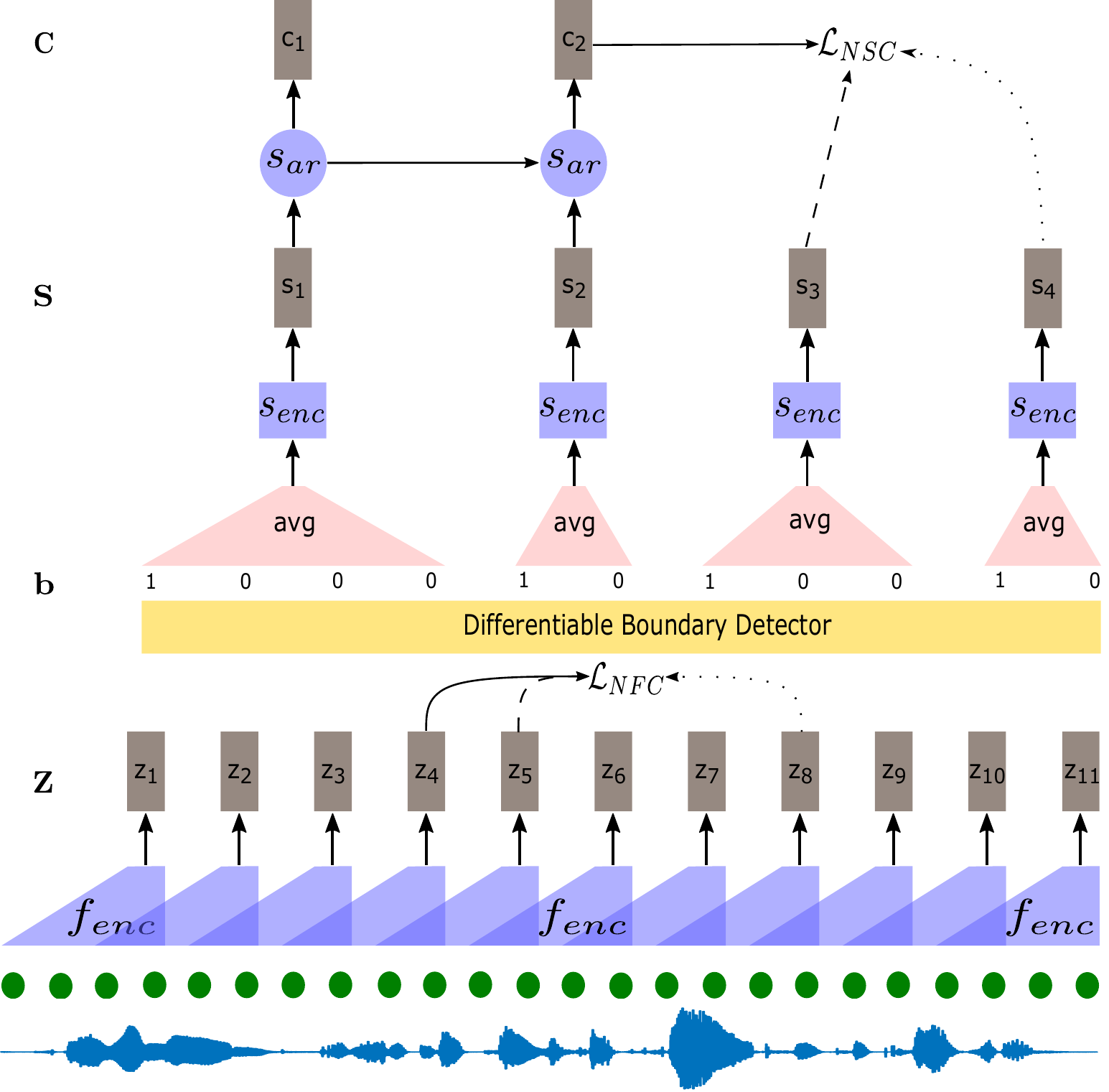}
    \vspace{-4mm}
    \caption{Overview of the Segmental Contrastive Predictive coding architecture. The solid line represents the reference frame (segment) $\zvi{4} (\cvi{2})$, the dashed line represents the positive  frame (segment) $\zvi{5} (\svi{3})$ and the dotted line represent the negative distractor randomly sampled from the signal}
    \label{fig:SCPC}
    \vspace{-6mm}
\end{figure}

We train the Segmental Contrastive Predictive Coding (SCPC) system, depicted in Figure~\ref{fig:SCPC}, by solving contrastive tasks at multiple scales: at the frame level, $ \mathcal{L}_{\mathrm{NFC}}$, and at the segment level, $\mathcal{L}_{\mathrm{NSC}}$, which require the model to identify the true latent representation within a set of distractors. The system's final objective is to minimize the overall loss ($ \mathcal{L}$), composed of both the next frame prediction and the next segment prediction losses:
\begin{equation}
    \mathcal{L} = \mathcal{L}_{\mathrm{NFC}} + \mathcal{L}_{\mathrm{NSC}} \;
\end{equation}

\subsection{Next Frame Classifier}

Let the sequence $\mathbf{X} = (x_1,x_2,...,x_T)$ represent a waveform. 
We learn an encoding function $f_{\mathrm{enc}}:\mathbf{X}\rightarrow\mathbf{Z}$ that maps audio sample sequences to latent spectral representations, $\mathbf{Z}(\in \mathbb{R}^{p\times L}) = (\zvi{1},\zvi{2},...,\zvi{L})$ at low frequency. Each p-dimensional vector $\zvi{i}$ corresponds to a 30 ms audio frame extracted with 10 ms shift. Given frame $\zvi{t}$, the model is trained to identify the next frame $\zvi{t+1}$ correctly within a set of $K+1$ candidate representations $\Tilde{\zvi{}} \in \mathcal{Z}_{t}$, which includes $\zvi{t+1}$ and $K$ distractors--randomly sampled from the same utterance, as

\begin{equation}
    \mathcal{L}_{\mathrm{NFC}} = -\log \frac{\exp(\mathrm{sim}(\zvi{t},\zvi{t+1}))}{\sum_{\Tilde{\zvi{}} \in \mathcal{Z}_{t} } \exp(\mathrm{sim}(\zvi{t},\Tilde{\zvi{}} ))}
\end{equation}
where $\mathrm{sim}(\mathbf{x},\mathbf{y}) =  \frac{\mathbf{x}\mathbf{y}^{T}}{\Vert \mathbf{x} \Vert \Vert \mathbf{y} \Vert} $ denotes the cosine similarity.

\subsection{Differentiable Boundary detection}
A phoneme segment can be thought of as a sequence of frames. The frames that follow each other with high similarity are likely to be from the same segment. If there is a high dissimilarity between adjacent frames, then it might indicate a segment change. $\mathbf{d}=(d_1,d_2,...,d_{L-1})$ captures the dissimilarity between two adjacent frames,
\begin{equation}
\begin{split}
& \mathbf{d} = 1 - (\mathbf{d^{s}} - \mathrm{min}(\mathbf{d^{s}})) / (\mathrm{max}(\mathbf{d^{s}}) -  \mathrm{min}(\mathbf{d^{s}})) \\
 \end{split}
\end{equation}
where $\mathbf{d^s} = (d^s_1,d^s_2,...,d^s_{L-1})$ s.t. $ d^{s}_{t} = \mathrm{sim}(\zvi{t},\zvi{t+1}) $  captures similarity between adjacent frames. To locate the segment boundaries, we need to find the frame indexes with high dissimilarity values. We can do that by finding peak locations in $\mathbf{d}$. Equation~\ref{eq:peaks} defines the peak detectors $\mathbf{p^{(1)}}$, $\mathbf{p^{(2)}}$ and $\mathbf{p}$ whose $t^{th}$ entries are
\begin{equation}
\label{eq:peaks}
\begin{split}
&    p_t^{(1)} = \min(\max(d_{t} -  d_{t-1},0), \max(d_{t} -  d_{t+1},0)) \\
&    p_t^{(2)} = \min(\max(d_{t} -  d_{t-2},0), \max(d_{t} -  d_{t+2},0)) \\
&    p_t = \min(\max(\max(p_t^{(1)},p_t^{(2)}) - \mathrm{thres},0), p_t^{(1)})
\end{split}    
\end{equation}
$p_t^{(1)}$ captures if $d_t$ is greater than both $d_{t-1}$ and $d_{t+1}$ or not. It will be zero if it is smaller than either $d_{t-1}$ or $d_{t+1}$ and non-zero otherwise. By itself, $p_t^{(1)}$ might be noisy and over-segment. We amend that by introducing $p_t^{(2)}$, which compares $d_t$ with $d_{t-2}$ and $d_{t+2}$. A peak in the low range can be more informative than one that is higher but otherwise is a trivial member of a tall range. So we check if the height of the peak as compared to its neighbors is more than a threshold or not instead of the peak height directly.

The vector $\mathbf{p}$ will have zeros wherever there are no boundaries and non-zero if there is a boundary. The non-zeros are values different across peaks, so we scale them to make all the non-zero values consistently $1$. We can do that by taking $\tanh$ of a large scalar multiple e.g. 100 of $p$, but this might result in vanishing gradients. We use a gradient straight-through estimator~\cite{oord2017neural,bengio2013estimating} 
for obtaining the boundary variables, 
\begin{equation}
\label{eq:bounds}
\begin{split}
&    \mathbf{b_\mathrm{soft}} = \tanh(10\;\mathbf{p}) \\
&    \mathbf{b_\mathrm{hard}} = \tanh(1000\;\mathbf{p}) \\
&    \mathbf{b} = \mathbf{b_\mathrm{soft}} + 
\mathrm{sg}(\mathbf{b_\mathrm{hard}}-\mathbf{b_\mathrm{soft}})
\end{split}
\end{equation}
where $\mathrm{sg}$ is the stopping gradient function that avoids gradients to flow through $\mathbf{b}_\mathrm{hard}$, which has exploding gradient at 0.
The $\mathbf{b}$ is $1$ at boundaries and $0$ elsewhere. This method can detect segments in a batch manner without any assumptions on segment length or the number of segments.

\begin{figure}
    \centering
    \includegraphics[width=1.05\columnwidth]{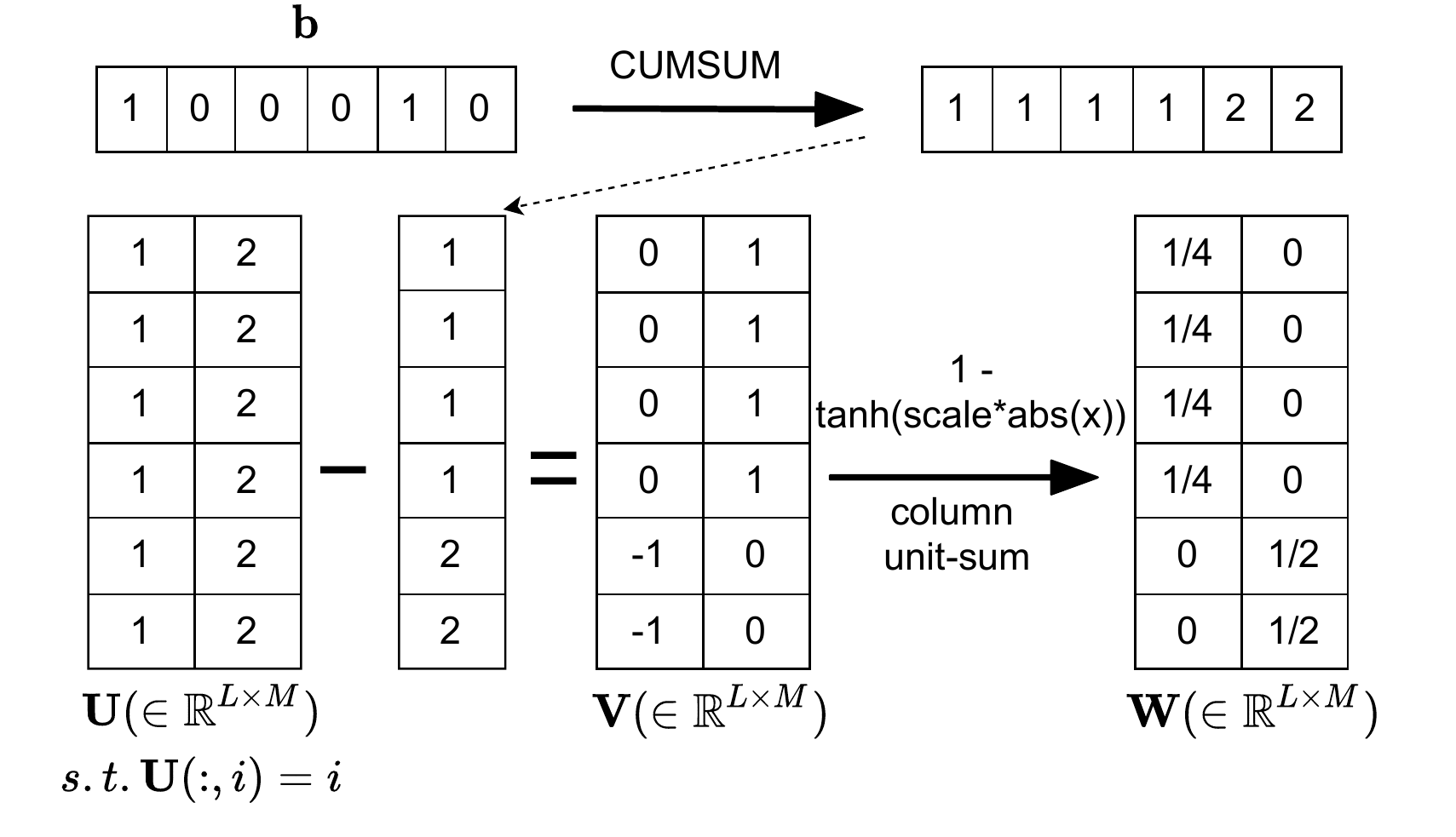}
    \caption{Weight generation for obtaining segment means in a differentiable manner. The number of segments $M$ can be computed by taking the sum of $\mathrm{bounds}$.}
    \label{fig:mu_gen}
    \vspace{-5mm}
\end{figure}

\begin{table*}[ht!]
\caption{Comparison of phoneme segmentation performance on TIMIT and Buckeye test sets. All the results use 20 ms tolerance window.}
\label{tab:phoneme}
\vspace{-2mm}
\centering
\begin{tabular}{@{}lllll|llll@{}}
\toprule

& \multicolumn{4}{c|}{TIMIT} & \multicolumn{4}{c}{Buckeye} \\
 \cmidrule(l){2-9}
 & Precision & Recall & F1 & \multicolumn{1}{l|}{R-val} & Precision & Recall & F1 & R-val \\ \midrule
RNN ~\cite{michel2016blind} & 74.80 & 81.90 & 78.20 & 80.10 & 69.34 & 65.14 & 67.18 & 72.13 \\
RNN Gate ~\cite{wang2017gate} &  &  &  & \multicolumn{1}{l|}{83.16} & 69.61 & 72.55 & 71.03 & 74.83 \\
CPC ~\cite{kreuk2020self} & 83.89 & 83.55 & 83.71 & \multicolumn{1}{l|}{86.02} & 75.78 & 76.86 & 76.31 & 79.69 \\
CPC+ ~\cite{kreuk2020self} & 84.11 & 84.17 & 84.13 & 86.40 & 74.92 & \textbf{79.41} & 77.09 & 79.82 \\
SCPC & \textbf{84.63} & \textbf{86.04} & \textbf{85.33} & \multicolumn{1}{l|}{\textbf{87.44}} & \textbf{76.53} & {78.72} & \textbf{77.61} & \textbf{80.72} \\ \bottomrule
\end{tabular}
\vspace{-4mm}
\end{table*}

\subsection{Segment representations}
After the boundary detection step, the feature sequence $\mathbf{Z} = (\zvi{1},\zvi{2},...,\zvi{L})$ is segmented into disjoint contiguous segments $\mathbf{S} = (\svi{1},\svi{2},...,\svi{M})$. 
The next step is to obtain representations for these variable-length segments. A fixed-size representation for segments would be easier to work with and would allow different segments to be compared straightforwardly. Since we decide a segment boundary based on high dissimilarity between frames, the frames that lie within a segment will be close to each other. We represent a segment via the average of constituting frames and feed the averages through a segment encoder $\mathrm{s_{\mathrm{enc}}}$ to generate the segment representations. From the Figure~\ref{fig:SCPC}, we see that the first four frames, $\zvi{1:4}$ belong to the first segment $\svi{1}$, and the next two frames $\zvi{5:6}$ belong to the second segment $\svi{2}$. The segment representations are given as $\svi{1} = \mathrm{s_{\mathrm{enc}}} (\frac{1}{4} \sum_{i=1}^{4} \zvi{i} )$ 
$\svi{2} = \mathrm{s_{\mathrm{enc}}} ( \frac{1}{2} \sum_{i=4}^{5} \zvi{i} ) $. 
We could iteratively compute those averages segment by segment but that would be prohibitively slow. Instead, we propose to vectorize those computations following the steps in Figure \ref{fig:mu_gen}. Segment representation can be obtained by multiplying the $\mathbf{Z}(\in \mathbb{R}^{p\times L})$ and $\mathbf{W}(\in \mathbb{R}^{L\times M})$ and feeding it through $\mathrm{s_{\mathrm{enc}}}$. 

\subsection{Next Segment Classifier}
The segment encoder $\mathrm{s_{\mathrm{enc}}}$ takes the segment averages as input and outputs the segment representations. We use a recurrent neural network (RNN), $s_\mathrm{ar}: \mathbf{S} \rightarrow \mathbf{C} $, to build a contextual representation $(\cvi{1},\cvi{2},...,\cvi{M})$ computed as $c_i = s_\mathrm{ar}(s_{i})$.
Given a reference representation $\cvi{t}$ the model needs to identify the next segment  $\svi{t+1}$ correctly from a set of $K+1$ candidate representations, $\Tilde{\svi{}} \in \mathcal{S}_{t}$ which includes $\svi{t+1}$ and $K$ distractors. 
\begin{equation}
    \mathcal{L}_{\mathrm{NSC}} = -\log \frac{\exp(\mathrm{sim}(\cvi{t},\svi{t+1}))}{\sum_{\Tilde{\svi{}} \in \mathcal{S}_{t} } \exp(\mathrm{sim}(\cvi{t},\Tilde{\svi{}} ))}
\end{equation}
where $\mathrm{sim}$ denotes the cosine similarity.

\subsection{Inference}
During inference, for a new utterance $\mathbf{X}$ we first extract the frame-level features, $\mathbf{Z}$, using $\mathrm{f_{\mathrm{enc}}}$.
For phoneme segmentation, the model outputs the dissimilarity between adjacent frames. The frames with high dissimilarity are considered segment boundary candidates.  Similar to~\cite{wang2017gate,michel2016blind,kreuk2020self}, we apply a peak detection algorithm to find the final segment boundaries. The peak prominence value for the peak detection algorithm is fined-tuned on the validation dataset~\cite{kreuk2020self}. For word segmentation, the model outputs dissimilarity score between context representation and the segment representation. This can be interpreted as how likely the model thinks $\svi{t}$ and $\svi{t+1}$ occur together. Segments that form a word are more likely to be occur together. A peak detection algorithm locates the times with lower prediction which are used as word boundary candidates.

\section{Experiments}

\subsection{Datasets and Evaluation Metrics}

We evaluate the proposed model on both TIMIT~\cite{garofolo1993timit} and Buckeye~\cite{pitt2005buckeye} datasets. For the TIMIT dataset, the standard train/test split is used, with $10\%$ of the train data randomly sampled employed as the validation subset. 
For Buckeye, we split the data into 80/10/10 for train, test, and validation sets, similarly to~\cite{kreuk2020self}. The longer recordings are divided into smaller ones by cutting during untranscribed fragments, noises and silences for ease of training. Each smaller sequence started and ended with a maximum of 20 ms of non-speech.
To make results more comparable to~\cite{kamper2020towards}, we also train our word segmentation system on the English training set from the ZeroSpeech 2019 Challenge~\cite{dunbar2019zero} and test on Buckeye. ZeroSpeech English set contains around 15 hours of speech from over 100 speakers. 
Using two different training and evaluation datasets allows us analyzing how well our method generalizes across corpora. 

For both the phoneme and word segmentation tasks, we measure the performance with precision (P), recall (R) and F-score with a tolerance of 20 ms~\cite{michel2016blind,wang2017gate,kreuk2020self}. F-score is not sensitive enough to capture the trade-off between recall and Over Segmentation (OS), defined as $R/P - 1$, i.e., even a segmentation model that predicts a boundary every 40 ms can achieve high F1-score by maximizing recall at the cost of low precision. This motivated a more robust metric, R-value~\cite{rasanen2009improved}, more sensitive to OS and the only way to obtain a perfect score (1) is to have perfect recall (1) and perfect OS (0). All the results reported here are an average of three different runs with different seeds unless stated otherwise.

\begin{figure*}[h!]
    \centering
    \subfloat[phoneme segmentation]{\includegraphics[width=1.7in]{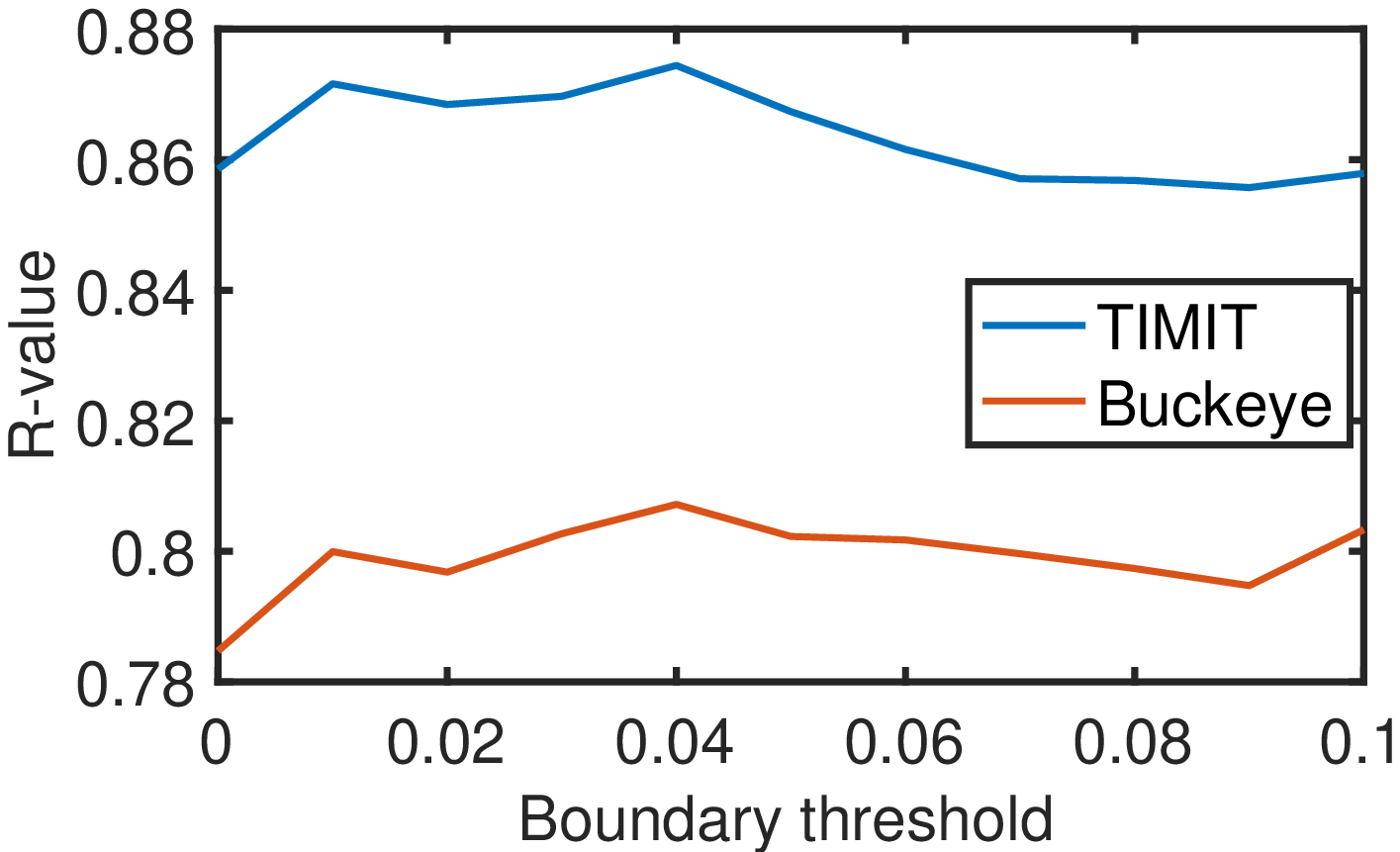}}
    \subfloat[word segmentation]{\includegraphics[width=1.7in]{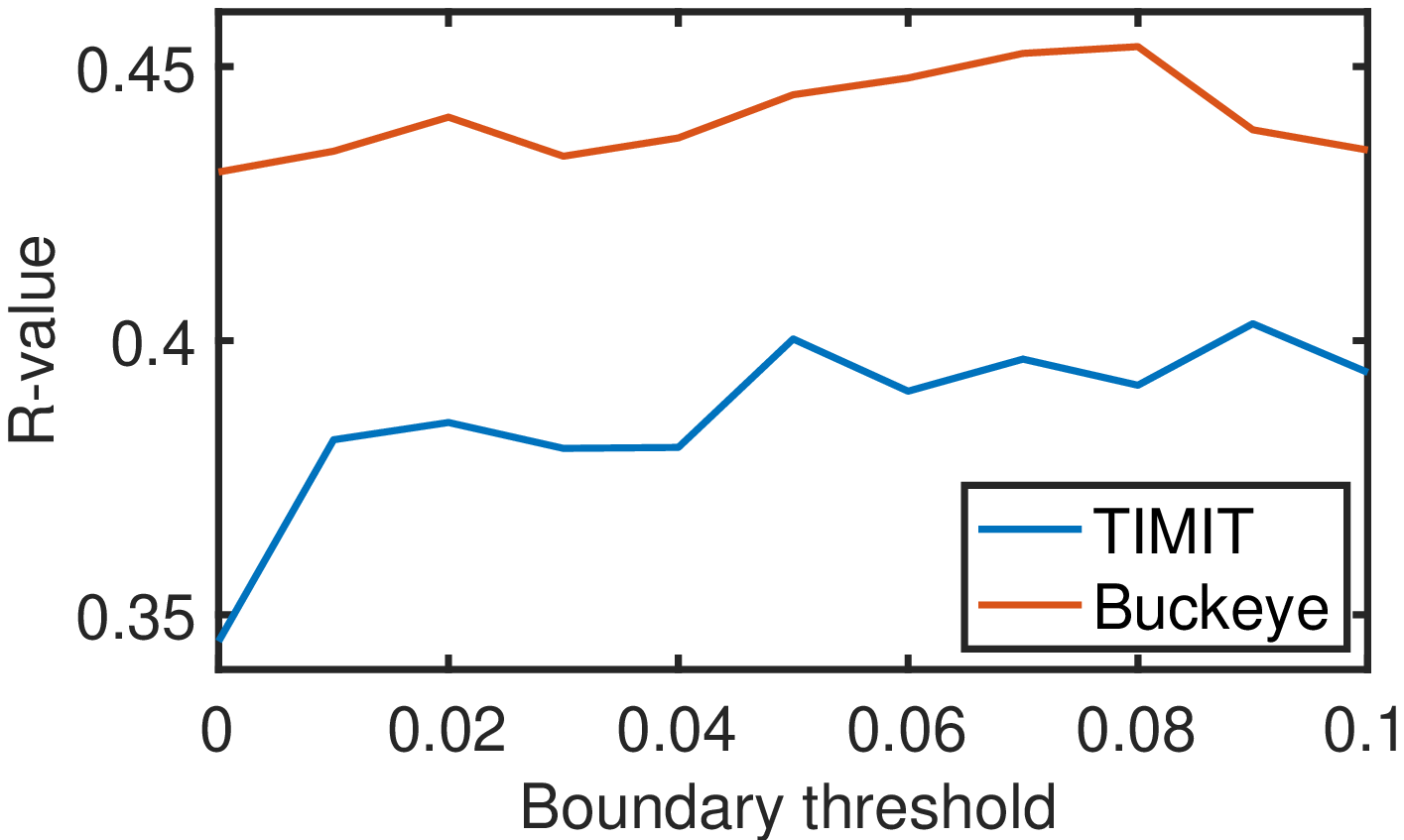}}
    \subfloat[phoneme segmentation]{\includegraphics[width=1.7in]{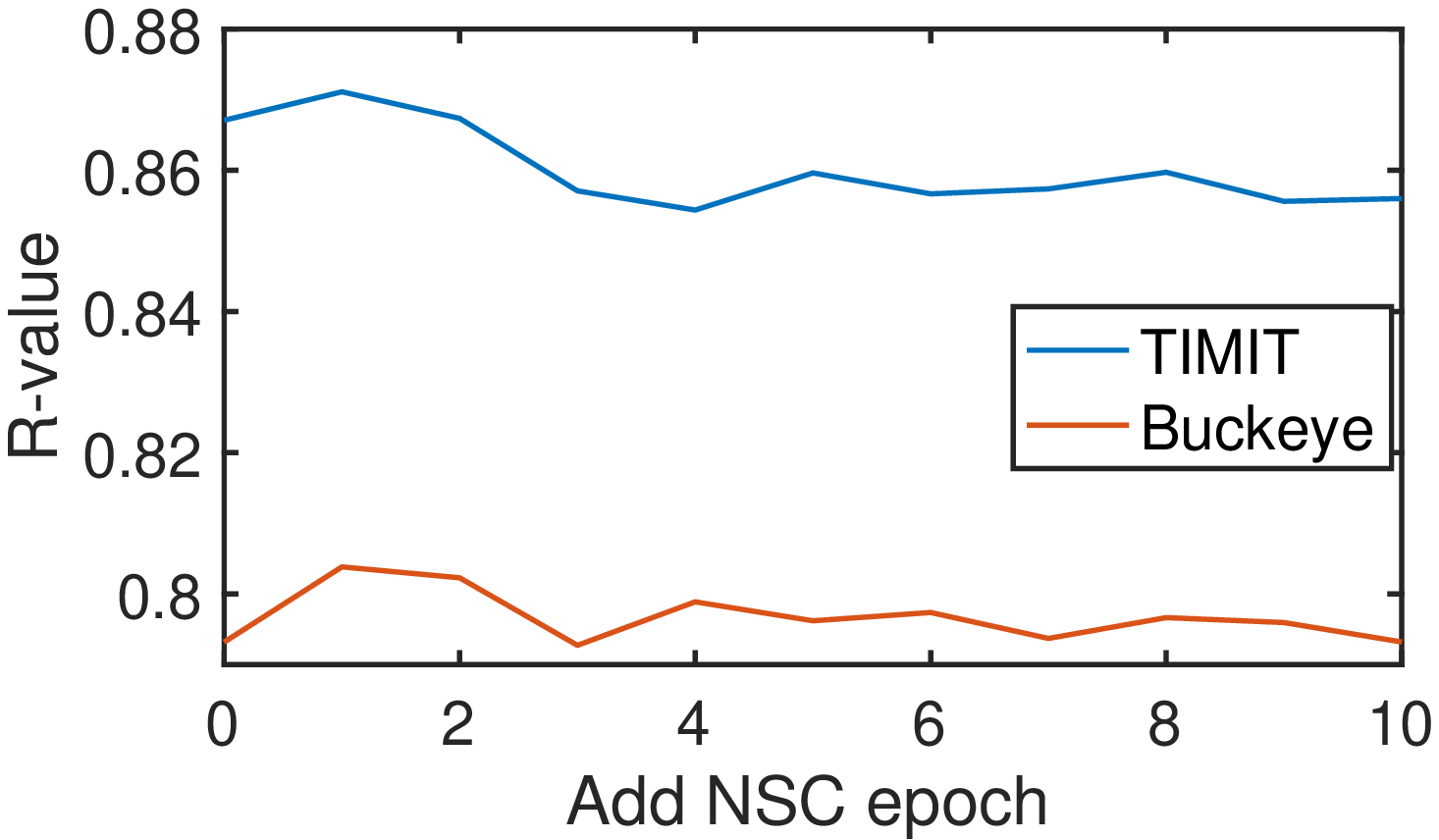}}
    \subfloat[word segmentation]{\includegraphics[width=1.7in]{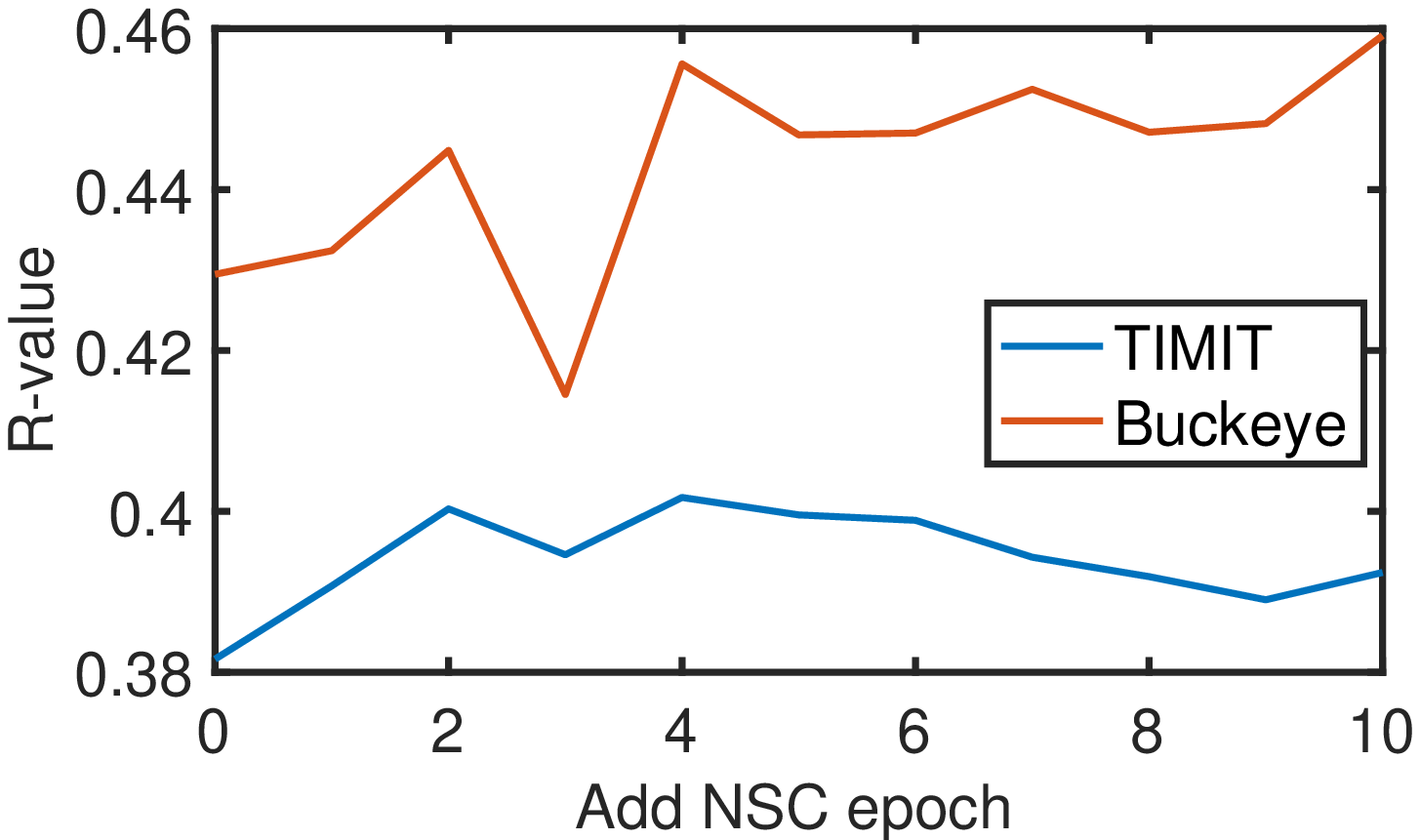}}
    \caption{Segmentation performance on test portions from TIMIT and Buckeye vs boundary threshold value and add nsc epoch}
    \vspace{-0.2in}
    \label{fig:R_valVsPara}
\end{figure*}

\subsection{Effect of Segmental CPC on phoneme segmentation}

As observed from Table \ref{tab:phoneme}, SCPC outperformed all the baselines. Our proposed approach, SCPC, extracts more structure from the speech data i.e. at frame-level and at phoneme-level, whereas the baselines rely on just the frame-level structure. 

The CPC+ line denotes the CPC system trained with additional unlabelled training data from Librispeech~\cite{kreuk2020self}. For TIMIT, the $100$ hours partition and for Buckeye $500$ hours partition are added to their respective training sets. SCPC outperforms the CPC trained with just the TIMIT and Buckeye datasets and with data augmentations from Librispeech.

\subsection{Word segmentation performance}
The word segmentation performance on the Buckeye dataset is shown in Table~\ref{tab:word}. SCPC clearly outperforms both the neural~\cite{kamper2020towards} and non-neural~\cite{kamper2017embedded,kamper2017segmental} methods for word segmentation. Unlike ES K-Means~\cite{kamper2017embedded} and BES GMM ~\cite{kamper2017segmental}, our proposed method does not require an initial segmentation method and can generate and adjust the boundaries during the learning process.
VQ-CPC and VQ-VAE are pretrained without the segmentation task~\cite{van2020vector} and then optimized for word segmentation, where SCPC is trained jointly. Our model is explicitly encouraged to assign blocks of feature frames to the same segment during the learning process. These approaches have different steps for feature extraction, feature learning, initial segmentation, and then word segmentation. In contrast, in our case, everything is done by a single model jointly with feedback from each other to improve performance.

Buckeye\_SCPC denotes the SCPC system trained on Buckeye dataset and tested on same, and 
ZS\_SCPC denotes SCPC system trained on Zerospeech 2019 dataset. 
The validation and test performance are very close, which shows that the model is generalizing well. 
The high performance across datasets shows the robustness of this approach. The slight difference in performance is because ZS\_SCPC is trained on more data. 

\begin{table}[]
\caption{Word boundary segmentation performance on Buckeye development dataset. Parenthesis show performance on test set. All the results use a 20 ms tolerance window. }
\vspace{-2mm}
\label{tab:word}
\resizebox{\columnwidth}{!}{
\begin{tabular}{@{}llllll@{}}
\toprule
Model & P & R & F1 & OS & R-value \\ 
\midrule
ES K-Means~\cite{kamper2017embedded} & 30.7 & 18.0 & 22.7 & -41.2 & 39.7 \\
BES GMM~\cite{kamper2017segmental} & 31.7 & 13.8 & 19.2 & -56.6 & 37.9 \\
VQ-CPC DP~\cite{kamper2020towards} & 15.5 & \textbf{81.0} & 26.1 & 421.4 & -266.6 \\
VQ-VAE DP~\cite{kamper2020towards} & 15.8 & 68.1 & 25.7 & 330.9 & -194.5 \\
AG VQ-CPC DP~\cite{kamper2020towards} & 18.2 & 54.1 & 27.3 & 196.4 & -86.5 \\
AG VQ-VAE DP~\cite{kamper2020towards} & 16.4 & 56.8 & 25.5 & 245.2 & -126.5 \\
ZS\_SCPC & \textbf{36.9} & 29.9 & \textbf{33.0} & -19.1 & \textbf{45.6} \\
& (34.8) & (31.0) & (32.8) & (-10.8) & (44.5) \\
Buckeye\_SCPC & 35.0 & 29.6 & 32.1 & \textbf{-15.4} & 44.5 \\ 
& (33.3) & (29.7) & (31.4) & (-10.8) & (43.4) \\ 
\bottomrule
\end{tabular}
}
\vspace{-4mm}
\end{table}

\subsection{Effect of boundary threshold}

To analyze the impact of boundary threshold on system performance, we vary the threshold from $0$ to $0.1$ with 0.01 step size and trained SCPC models. The segmental loss is added after two epochs. As observed in the Figure~\ref{fig:R_valVsPara}(a,b) the phoneme segmentation performs better with lower thresholds. Both datasets achieve the best phoneme segmentation performance with a peak threshold of $0.04$. The optimal threshold for the word segmentation is higher around $0.09$ and $0.08$ for TIMIT and Buckeye respectively. We hypothesize that a lower peak threshold allows the model to generate more boundaries which can match better with the higher number of ground truth phoneme boundaries present where word boundaries are fewer, so a higher peak threshold suppresses the unnecessary boundaries for improved performance. 

\subsection{When to add NSC loss}
To analyze the importance of the epoch at which the segmental loss is added to the model objective, we trained SCPC where the segmental loss is added after $i^{th}$ epoch and $i$ is varied from $0$ to $10$. The boundary threshold is kept at $0.05$. As observed, in Figure~\ref{fig:R_valVsPara}(c,d) the phoneme segmentation performs better if segmental loss is added early ($i = 1$) and word segmentation performs better when segmental loss is added late ($i = 4$). We hypothesize that this allows the model to learn better frame-level features before combining them into segments. 

\section{Conclusions} 

This work proposes methods to exploit speech signal structure at the segment level and extend the self-supervised learning framework beyond frame-level auxiliary tasks. We propose a differentiable boundary detector to find variable-length segments. Our experimental results indicate that the learned segments correspond to phoneme-like units. Our model is trained to jointly optimize frame and segment level representations from raw speech waveform. Our model outperforms existing phoneme and word segmentation methods on TIMIT and Buckeye datasets.

\bibliographystyle{ieeetr}
\bibliography{ref} 

\end{document}